\documentclass[11pt]{article}
\usepackage[margin=3cm]{geometry}
\usepackage[utf8]{inputenc}
\usepackage[english]{babel}
\usepackage[T1]{fontenc}
\usepackage{float}
\usepackage{authblk}
\usepackage{amsmath}
\usepackage{graphicx}
\usepackage[colorinlistoftodos]{todonotes}
\usepackage[colorlinks=true, allcolors=blue]{hyperref}
\usepackage{amsmath}
\usepackage{color}
\usepackage{xcolor}
\usepackage[super]{nth}
\date{}

\title{Solitons, dispersive shock waves and Noel Fredrick Smyth}
\author[1*]{Saleh Baqer}
\author[2,3]{Tim Marchant}
\author[4]{Gaetano Assanto}
\author[5]{Theodoros Horikis}
\author[6]{Dimitri Frantzeskakis}
\affil[1]{Department of Mathematics, Kuwait University, Kuwait City 13060; email: saleh.baqer@ku.edu.kw}
\affil[2]{School of Mathematics and Applied Statistics, University of Wollongong, 2522 New South Wales, Australia; email: tim@uow.edu.au}
\affil[3]{Australian Mathematical Sciences Institute, University of Melbourne, Melbourne, VIC 3052, Australia}
\affil[4]{NooEL—Nonlinear Optics \& OptoElectronics Laboratory, University of Rome `Roma Tre', Rome 00146, Italy; email: assanto@uniroma3.it}
\affil[5]{Department of Mathematics, University of Ioannina, Ioannina 451 10, Greece; email: horikis@uoi.gr}
\affil[6]{Department of Physics, National and Kapodistrian University of Athens, Athens 15 784, Greece; email:
dfrantz@phys.uoa.gr}

\begin{document}

\maketitle

Noel Frederick Smyth (NFS), a Fellow of the Australian Mathematical Society and a Professor of Nonlinear Waves in the School of Mathematics at the University of Edinburgh, passed away on February 5, 2023. NFS was a prominent figure among applied mathematicians who worked on nonlinear wave theory in a broad range of areas. Throughout his academic career, which spanned nearly forty years, NFS developed mathematical models, ideas, and techniques that have had a large impact on the understanding of wave motion in diverse media. His major research emphasis primarily involved the propagation of solitary waves, or solitons, and dispersive shock waves, or undular bores, in various media, including optical fibers, liquid crystals, shallow waters and atmosphere. Several approaches he developed have proven effective in analyzing the dynamics and modulations of related wave phenomena. This tribute in  the journal of \textit{Wave Motion} aims to provide a brief biographical sketch of NFS, discuss his major research achievements, showcase his scientific competence, untiring mentorship and unwavering dedication, as well as share final thoughts from his former students, colleagues, friends, and family. The authors had a special connection with NFS on both on personal and professional levels and hold deep gratitude for him and his invaluable work. In recognition of his achievements in applied mathematics, \textit{Wave Motion} hosts a Special Issue entitled ``Modelling Nonlinear Wave Phenomena: From Theory to Applications,'' which presents the recent advancements in this field.

\begin{figure}[!htp]
    \centering
    \includegraphics[width=0.5\textwidth]{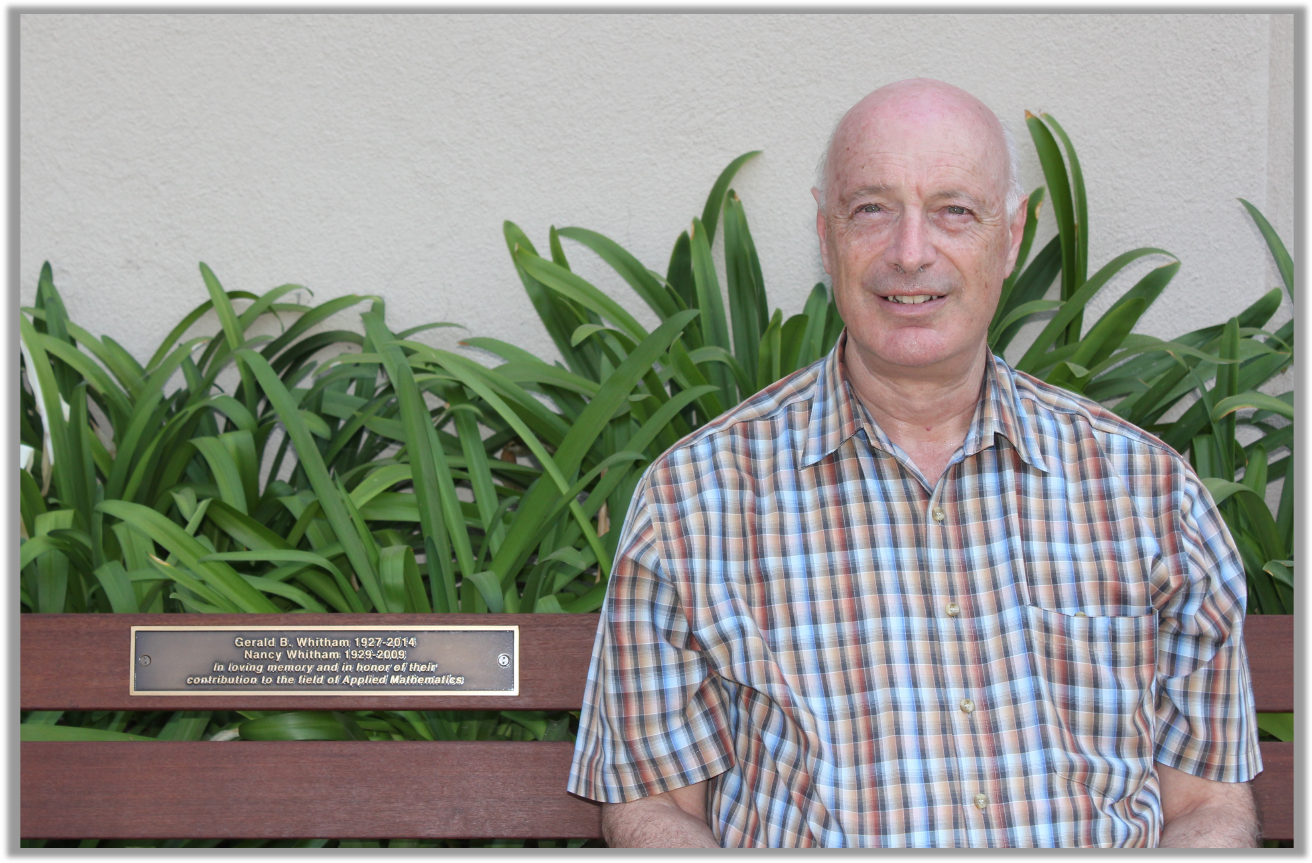}
    \caption{Professor Smyth at the Memorial Bench of his PhD advisor Professor Gerald B. Whitham (FRS) at the California Institute of Technology (Caltech) in August 2015.}
    \label{f:bench}
\end{figure}


\section{\centering\sc\textcolor{purple}{Researcher, Teacher and Mentor}}

Noel Frederick Smyth (NFS) was a distinguished researcher in the mathematics applied to nonlinear physics, specifically nonlinear dispersive wave propagation in fluids and optics. Unlike  mathematical research pursued for academic purposes, NFS' research was focused on the development and use of  models and tools with direct impact on applications. His interest was in deriving mathematics that could describe physical phenomena and comparing theories with experimental results or numerical simulations for scientific validation. He was particularly interested in solitary waves (or solitons) and dispersive shock waves that occur in various natural and technological systems. To study these nonlinear wave phenomena, he utilized partial differential equations, asymptotic analysis, perturbation methods, numerical approaches, variational calculus, and Whitham modulation theory \cite{whitham, whitham1, whithampert, whithamvar}.

NFS published over 150 papers and five book chapters with researchers from different scientific backgrounds, working with scientists from various countries such as Australia, the USA, the UK, Italy, France, Portugal, Germany, Mexico, Switzerland, Israel, Greece, Turkey, Poland and Kuwait. He was particularly proud of his long-term and productive collaborations with experimental groups in nonlinear optics in Rome (Professor Gaetano Assanto) and Warsaw (Professor Miroslaw Karpierz). His work on mathematics applied to physics earned him nine Editors' Picks Awards from the American Physical Society (APS) and the Optical Society of America (OSA). He was appointed Senior Member of the OSA in 2011, Fellow of the Australian Mathematical Society (FAustMS) in 2015 and received the ``Outstanding Referee'' award in 2016 from the prominent American Physical Society's journals \textit{Physical Reviews} and \textit{Physical Review Letters}.

Aside from his research, Prof Smyth was an excellent teacher who gave various applied mathematics courses at the University of Edinburgh. He was well known for his competence  and teaching style that did not involve copying lecture notes on blackboards. His open-door policy for office hours was a highlight among his students, allowing them to seek his guidance and support whenever they had questions. He supervised and mentored several students and researchers, successfully supervising five MS and eleven PhD students on water waves and nonlinear optics projects at the University of Edinburgh, as well as five PhD students at the University of Wollongong. His strategy in choosing research problems for prospective students was unique. Namely, he tackled  open scientific questions by himself, and  only if they appeared promising for a successful project he would propose them to the students.

NFS was a caring, patient and supportive mentor who tirelessly guided his students until they fully grasped the subject they worked on and became independent. He would take his time and answer all students' questions satisfactorily. Indeed, he enjoyed conversing for hours with them and fostering a warm and supportive environment for their scientific growth and development. His office was a platform for intellectual enrichment. His students were able to get good job positions in industry and academia after graduating, a testament to his excellent mentorship, as described in Section \ref{s:thoughts1} below.  

\begin{figure}[!htp]
    \centering
    \includegraphics[width=0.5\textwidth]{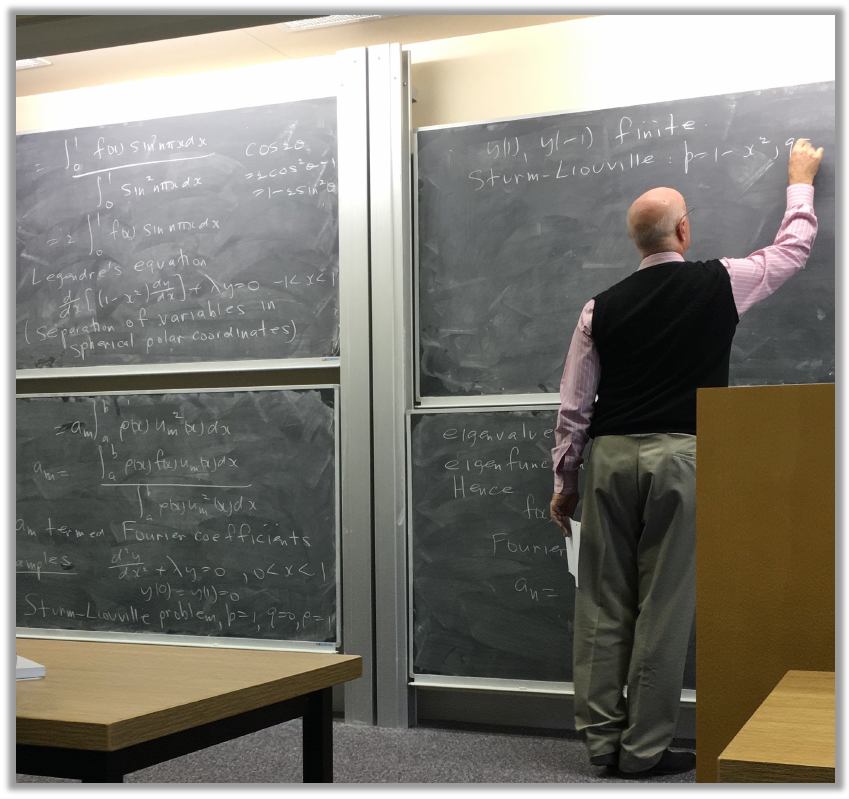}
    \caption{NFS giving a lecture on Advanced Methods of Applied Mathematics (Honours Class) in the Department of Mathematics at the University of Edinburgh, King's Buildings, October 2016.}
\end{figure}

\section{\centering\sc\textcolor{purple}{Early Life and Education}}

NFS was born on May 16, 1958, in Queensland, Australia, and spent his childhood there. His mother's family had settled in central Queensland in the 1840s, and he remained a proud Queenslander all his life.

He began his education at Salisbury State School from 1964 to 1967 before moving on to Jindalee State School from 1968 to 1970. He then attended Toowong State High School, where he graduated in 1975. His final year at high school was especially successful, as he received several prizes for his  achievements: he was awarded the Mathematics Prize and the Science Prize as the top student in these subjects in the school that year. The highest accolade he received was the Dux Prize, which is the most prestigious award given to a student in a given year by an Australian school. The high school headmaster informed NFS that he had ranked first in the state of Queensland.

After completing high school, NFS enrolled in Chemistry major at the University of Queensland. However, he shifted his focus to Mathematics major during his second year after finding out that he was much more interested in mathematics than chemistry. 
His favorite subjects were in applied mathematics, particularly those related to waves and fluids. 
He earned an Honors Degree and conducted his research under the supervision of Professor A.F. Pillow. The title of his project was ``\textit{Survey of Hydrodynamic Stability with Particular Emphasis on Linear and Nonlinear Stability of Couette Flow Between Rotating Cylinders.}'' During his studies at the University of Queensland, he received the Maude Walker Prize for the best first-year science student, the Priest Memorial Prize for the best second-year applied mathematics student, and the University Medal for an outstanding Honors Degree.

Upon completing his Honors Degree, NFS applied to the California Institute of Technology (Caltech) to pursue a PhD in Applied Mathematics. Interestingly enough, during his honors year Professor G.B. Whitham \cite{whithambio1, whithambio2} was on sabbatical at the University of Queensland. NFS was already familiar with Whitham's research and had already developed an interest in his work, especially his  book ``\textit{Linear and Nonlinear Waves}'', now a classic. This prior acquaintance and knowledge allowed him to engage in intellectually stimulating conversations with Whitham. 
Caltech eventually offered him 
a comprehensive scholarship for his PhD study and a Special Institute Fellowship.

NFS arrived at Caltech in 1980 and began working with Gerald B. Whitham, who he deemed to be the most suitable supervisor and mentor for his research in nonlinear wave theory. Along with taking different applied math courses, he worked on two research problems. One involved investigating the motion of a soliton on a slowly varying beach; for this he had the opportunity to collaborate with the Civil Engineering Group of Professor Fredric Raichlen. The other project dealt with the slowly varying modulated motion of capillary waves on a fluid sheet using the averaged Lagrangians, also known as Whitham modulation theory. He successfully completed both projects and titled his PhD thesis ``\textit{Part I. Soliton on a Beach and Related Problems. Part II. Modulated Capillary Waves}''\cite{noelthesis}. 

NFS had great respect for American people and the USA and was proud of his Caltech education. 
He traveled extensively throughout North America, visiting several western states and major national parks, including Yosemite, Zion, Bryce, Yellowstone, Grand Teton, Banff, and Jasper. He returned to the USA and Caltech multiple times after completing his PhD degree. His last visit  was when he delivered an address at the Whitham Memorial Bench Dedication on August 23, 2015 (see Fig. \ref{f:bench}).  

\section{\centering\sc\textcolor{purple}{Academic Career: from Caltech to Edinburgh}}

After NFS completed his PhD, Professor Whitham arranged for him to take on a post-doctoral position at Caltech and continuing his work on nonlinear dispersive wave theory. The post-doc focused on shock dynamics, as Whitham intended to use numerical and theoretical solutions of the gas dynamic equations to extend the range of what could be examined and compared with experimental results \cite{henshaw}. This was an interdisciplinary project, carried out in conjunction with the experimental aeronautical group of Professor Bradford Sturtevant (GALCIT, Graduate Aeronautical Laboratory in Caltech).

Towards the end of his academic time at Caltech, NFS was invited to undertake two post-doctoral positions, one from Professor Alan C. Newell at the University of Arizona and the other from Professor Robert M. Mirua at the University of British Columbia. However, NFS 
wanted to return to his home country, Australia. Once back in Australia in 1984, he began post-doctoral work under the supervision of Professor Roger Grimshaw at the University of Melbourne and the University of New South Wales, where Grimshaw had been offered a professorship position. His post-doctoral research at this time focused on resonant flow of stratified fluid over topography \cite{noelres1, noelres2}. 

In 1987, NFS took a lectureship position at the University of Wollongong and received a research grant from the Australian Research Council, which he used to advertise a research fellowship. Doctor Tim Marchant applied and henceforth their joint research centered on the Korteweg-de Vries (KdV) equation and its applications to fluids. One project involved deriving a higher-order KdV equation (known as the extended KdV equation) from the full Euler shallow water wave equations using multiple scales perturbation theory and variational calculus, as well as using Whitham modulation theory to find dispersive shock wave solutions \cite{ekdv}. The other project focused on  exact and approximate solutions of the initial-boundary value problem for the KdV equation on the semi-infinite line \cite{timkdvboundary}.

The relationship between Tim Marchant and NFS was more than just that of a mentor and a mentee, as it blossomed into a friendship that spanned over three decades. They kept collaborating and supervising research students on projects related to water waves and nonlinear optics, while NFS  visited Wollongong yearly to interact with Tim.

Advancing in his career, in 1990 NFS decided to move to the University of Edinburgh. 
Moving to Edinburgh opened NFS to more research opportunities and possibilities of collaborations  in other countries 
including Professors Tim Minzoni, Gaetano Assanto, Roger Grimshaw, Gennady El, and many more. 
He made an effort to visit his academic friends and collaborators every year, especially the ones in Mexico, Italy and Australia.

In Mexico, NFS visited the Department of Applied Mathematics and Mechanics of the National Autonomous University of Mexico (UNAM) for two to three months each year between 1995 and 2017,  collaborating with Professors A.A. Minzoni \cite{minzonitribute} and Gilberto Flores. Incidentally, and interestingly enough, Minzoni's  father in the Italian artillery and NFS' grandfather in the infantry had happened to fight on opposite sides in North Africa during the Second World War.  NFS had a great fondness for Mexico and its people, describing them as friendly, warm, and family-oriented. He spoke Spanish fluently and enjoyed traveling all over the country. All of Smyth's research trips to Mexico were financially covered by Mexican research grants (CONACYT), which he highly appreciated. 

After establishing contact and starting with Minzoni a fruitful collaboration and eventually friendship with Professor Gaetano Assanto in Rome (Italy), NFS engaged in research on nematicons and then traveled to Italy for over fifteen years to visit Assanto's team (NooEL) at the University ``Roma Tre.'' Such extended collaborations included the joint attendance of numerous conferences and meetings in Scotland, Italy, Mexico and Australia, including  research interactions with Poland (Warsaw Technical University) and India (Sastra University).  NFS and Assanto started and organized a series of international workshops on ``Nonlinear Guided Waves,'' which became an annual/biannual opportunity to meet with a network of researchers and the relevant community in a variety of destinations including Mexico, Australia and Morocco.  

NFS continued his dedicated work as a Professor of Nonlinear Waves in the School of Mathematics at the University of Edinburgh until, in 2021, he was diagnosed with gastroesophageal cancer. A year later, he received the devastating news that the cancer had spread to his brain. The illness  made it impossible for him to continue working and communicating as he had done before.

His illness came as a profound shock and tragedy to his family, friends, and collaborators. The National Health Service staff assigned to him were deeply impressed by his remarkable resilience in the face of such a formidable adversary within. Despite grim expectations, he fought the cancer with unwavering determination, earning the admiration of those who cared for him and keeping in contact online with the closest friends, including Tim Marchant and Gaetano Assanto.

Regrettably, on February 5, 2023, NFS lost his courageous battle with cancer. He is survived by his wife, Juliet, and their son, Calum. His passing has left a profound void within the community of applied nonlinear wave researchers, and his absence is deeply felt. If nonlinear wave equations were a language, NFS spoke it fluently. He will be dearly missed, and his contributions will be remembered with great reverence.

\begin{figure}[!htp]
    \centering
    \includegraphics[width=0.5\textwidth]{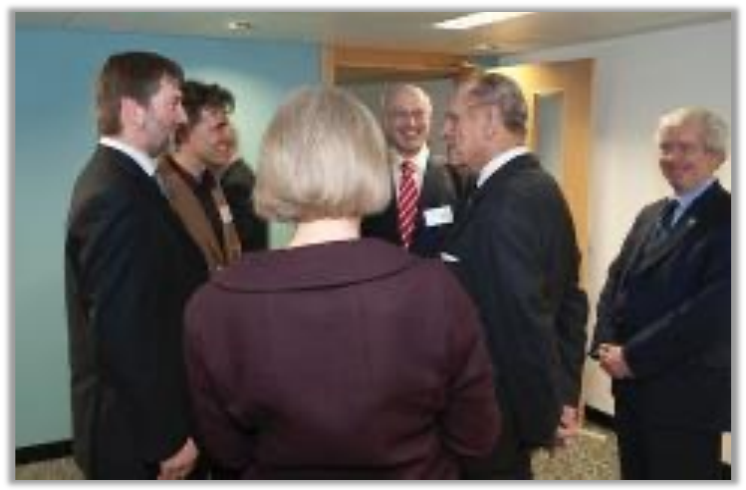}
    \caption{NFS, wearing a red tie on his promotion to Professor of Nonlinear Waves in 2009, together with other colleagues meeting the Duke of Edinburgh, Prince Phillip.}
    \label{f:duke}
\end{figure}

\section{\centering\sc\textcolor{purple}{Snapshots of NSF's Nonlinear Wave Research Interests}} 

Professor Smyth's research interests were primarily in wave motion and nonlinear waves in various media. He developed mathematical techniques to unravel quantitatively the mechanisms involved, interpreting and understanding the experimental observations and making predictions of novel phenomena. While this research work was conducted in the context of wave motion, it finds applicability to various areas of science, ranging from  nonlinear optics to water waves, from microwave heating to heat flow and diffusion, phase changes, oceanography, meteorology, geophysics, lasers, shock waves, particle physics, black holes, biophysics and epidemiology. Much of this endeavors involved interdisciplinary collaboration with scientists in other areas, particularly when modelling experiments in nonlinear optics. 

 \subsection{\centering\sc\textcolor{purple}{Nonlinear Dispersive Wave Equations and Solitons}}

The majority of NFS' research was on analyzing and finding solutions of nonlinear dispersive wave equations, in particular solitary wave and undular bore (dispersive shock wave) solutions. Examples of such equations are the Korteweg-de Vries, nonlinear Schr\"odinger, Benjamin-Ono, non-local Whitham and Sine-Gordon equations.  These equations arise in a wide range of application areas, including fluid mechanics, oceanography, meteorology, optics, solid mechanics, Bose-Einstein condensates, plasmas and  biology. Hence, any techniques and solutions developed for one application have implications across a wide area of science. The major techniques he used to analyze nonlinear dispersive wave equations are asymptotic theory, Whitham modulation theory and its extensions and numerical solutions.  The analytical and asymptotic techniques NFS and his collaborators developed have been used to model waves in nature, including waves in the ocean and the atmosphere, including resonant flow over topography, and light waves in nonlinear optical fibres, liquid crystals and colloids.
     
A major theoretical problem in nonlinear wave theory is the analytical study of the interaction of coherent nonlinear structures and linear radiation.  It is difficult to determine this interaction using the usual techniques of perturbed inverse scattering and soliton perturbation theory. NFS and his collaborators developed mathematical methods to study this interaction on a case by case basis within general classes of nonlinear systems, since each system has its
own qualitatively and quantitatively different behaviour.  Solutions obtained using these new techniques have been found to be in excellent agreement with full numerical solutions of the appropriate governing equations and have enabled the comprehension of the underlying mechanisms which result in the bi-directional coupling between the coherent structure and the radiation \cite{kathrad1, kathrad2, kathrad3}.  The approximate method has been successfully extended to study optical devices based on optical fibres, beam propagation in liquid crystals, defect formation in nano-electrical devices and energy propagation in protein and DNA chains, among others. In all these applications the radiation drives the evolution.

NFS had an ongoing research interest in the KdV-type equations and other universal nonlinear dispersive wave equations, such as the NLS-type equations. KdV-type equations arise in a large range of applications, particularly the propagation of waves in the ocean and atmosphere.  The evolution of waves governed by any of these nonlinear dispersive wave equations involves the shedding of dispersive radiation.  As for NLS-type equations, the interaction between  radiation and  evolving waves is extremely difficult to analyse using standard techniques. The asymptotic techniques developed by NFS and collaborators to analyse the interaction between solitary waves and shed radiation provided solutions in excellent agreement with full numerical solutions and observational results for atmospheric and oceanographic waves. NFS studied a wide range of waves governed by KdV-type equations, including waves forced by topographic features, waves in magma \cite{magma} and waves on the surface of a fluid. Physical examples he studied include the Morning Glory of the Gulf of Carpentaria and the internal tide \cite{morning2}. Jointly with Professor Grimshaw he developed the mathematical theory and solution for transcritical (resonant) flow over topography, based on Whitham modulation theory. The latter has now become a significant research area with applications to a wide range of physical flows.

 \subsection{\centering\sc\textcolor{purple}{Nonlinear Self-Guided Beams in Soft Matter}}


One of NFS' major areas of scientific involvement was the  formation and propagation of light beams in  nonlinear optical materials, thermal media and soft matter, from birefringent liquid crystals to colloids. This theoretical endeavor also involved Professors T. Minzoni, T. Marchant and A.L. Worthy.  Optical solitons in reorientational nematic liquid crystals - termed nematicons - were studied experimentally by NFS' collaborators, Professor Gaetano Assanto at the University ``Roma Tre'' \cite{JSTQE, physdrev, JNOPM} and Professor Miroslaw Karpierz at the Warsaw Technical University. \\
Nematic liquid crystals (NLCs) possess a ``huge'' effective nonlinear response orders of magnitude larger than in glass and many nonlinear dielectrics. Being birefringent, electrically tunable, transparent, saturable and nonlocal, NLCs are an ideal platform for the investigation of nonlinear optical effects, both experimentally and theoretically, towards potential low-power and compact devices in photonics. Based on the coupled system of nonlinear Scr\"odinger-type equation(s) for the  beam(s) evolution and an elliptic equation for the reorientational response, NSF derived approximate solutions of this non-integrable dispersive system subject to nonlocality, essentially based on Whitham's modulation theory and the interaction of shed radiation with self-trapped light beams \cite{Macneil, hinonl}). He developed models, theory and numerics for individual and multiple interacting single nematicons \cite{Minzoni, Panos}, vortex and vector solitary beams with coherent/incoherent interactions \cite{2color, 3color, antiguide, vortexnematicon, vectorvortex, vortexvector, Yana}, refraction and reflections of nematicons at dielectric interfaces or index perturbations \cite{steering, refraction, refra, scattering} among others. The work on these solutions in liquid crystals was also extended to dark and grey solitons \cite{dark, coupleddark}, discrete solitons in periodic waveguide arrays \cite{PRL}, nematicons in media with competing nonlinearities and/or in the presence of thermal effects \cite{temperature}, multi-hump solutions \cite{volcano}, curved solitary waves in non-uniform NLC samples \cite{Sala, bending, curving}, self-confined waves via spin-orbit interactions of light (i.e., geometric phase) \cite{spin}. NFS's physical insights always allowed him to combine mathematical techniques often based on Whitham's approach and averaged Lagrangians with conservation laws (energy, momentum), in order to get to the heart of the given problem without redundant details or cumbersome assumptions. 
NFS also extended his work towards the control of solitons via external voltages or potentials \cite{tunneling}, other beams \cite{extlight} and boundaries \cite{boundary}, aiming at reliable design tools for optical devices based on reorientational nematic liquid crystals. One of his last publications dealt with optical isolation using nematicon waveguides in samples with graded orientation of the molecular director \cite{OL22}.

\subsection{\centering\sc\textcolor{purple}{Dispersive Shock Waves, Modulation Theory and Non-convex Dispersion Effects}}

In addition to solitary waves, nonlinear media can generate a generic class of waves known as undular bores, dispersive shock waves (DSW), or collisionless shock waves. These terms essentially describe the same phenomena, with the first two being more commonly used in the context of waves in fluids \cite{hydraulic, stratified, dambreak}, meteorology \cite{loweratmosphere, morning1, morning2} and optics \cite{nonlocaldsws, superfluid,photoref, fibers}, while the third is prevalent in plasma physics \cite{plasma1, plasma2, plasma3, plasma4} and Bose-Einstein condensates \cite{bec1, bec2}. They are called dispersive as they are the dispersive analogue of classical (viscous) shock waves. A DSW arises from steep gradients and comprises a train of waves with solitary waves at one edge and small-amplitude (linear and harmonic) dispersive waves at the trailing edge. As a DSW propagates, it smooths out the corresponding mean level change.

Physical examples of Dispersive Shock Waves (DSWs) include tidal bores, tsunamis, and the cloud phenomena known as morning glory clouds. The theory of DSWs is well-developed in local media, such as fluids and optical fibers, where DSWs take classical forms. This means that the stationary levels ahead and behind the shock are connected by a well-defined modulated cnoidal wave, with the trailing edge being a linear harmonic wave and the leading edge being a solitary wave.

The underlying method for theoretically understanding the evolution and determining the macroscopic and microscopic properties of DSWs is Whitham modulation theory. This method is based on multiple scales perturbation theory \cite{whitham1} or, equivalently, averaged Lagrangians \cite{whithampert, whithamvar}. Modulation theory yields a system of Partial Differential Equations (PDEs) that govern the slowly varying wave parameters corresponding to the DSW, such as its wavenumber, mean level, and amplitude. If this system is hyperbolic, then the underlying wavetrain is modulationally stable, while if it is elliptic, the wavetrain is unstable \cite{whitham}.

Hyperbolic modulation systems give rise to DSW solutions, as found in \cite{plasmapit}. However, not all nonlinear, dispersive wave systems can be expressed in Riemann invariants, especially the non-integrable ones. Such expression is always possible and a guarantee for nonlinear, dispersive wave equations that are integrable, see \cite{analysis} for detailed discussion. Therefore, a challenge in the field has been to study DSWs in the absence of modulation equations, namely, without any knowledge of Riemann invariants and simple wave solutions. This led to a significant contribution by Professor Gennady El in developing a theoretical method that determines the macroscopic properties of DSWs, known as the DSW fitting method \cite{fitting}. The significance of this method lies in its ability to determine the velocities and amplitudes of both the trailing harmonic and leading solitary wave edges solely through the knowledge of the linear dispersion relation associated with the model, and therefore, no knowledge of the full modulation equations is required. For an in-depth discussion of DSWs and the fitting method, their practical applications and the relationship between Whitham modulation equations and DSWs, refer to \cite{dswreview}.

NFS applied, developed, and extended various asymptotic methods based on Whitham modulation theory to a wide range of dispersive hydrodynamic models (models formulated by conservation laws with dispersive corrections) encountered in real-world applications. These models include the Korteweg-de Vries (KdV) equation \cite{timkdvboundary, kdv1,kdv2,kpnoel} and its higher-order counterparts, such as the Kawahara equation \cite{pat, patkawahara, salehekdv}, the modified KdV equation \cite{mkdv} and the extended KdV equation \cite{ekdv,salehekdv, ekdhighermodu,ekdvinteract}. He also worked on KdV-type and forced KdV equations \cite{albalwi}, the Whitham equation \cite{xin}, the Sine-Gordon equation \cite{sine1, sine2, sine3, sine4, sine5}, the NLS-type equations in reorientational nematic liquid crystals \cite{physdrev, 2color, tunneling, modulnematic, mdpinembook, excellent}, the colloidal equations \cite{colloid1, colloid2, colloid3}, the Whitham-Boussinesq system of equations \cite{whithbous}, the Camassa-Holm equation \cite{camassa} and the fully nonlinear shallow water Euler wave equations \cite{whithbous, fullwater}. Comparisons in his studies with full numerical solutions of the associated models demonstrated that Whitham modulation theory gives excellent predictions. Another major achievement by NFS and collaborators was recognizing that the DSW fitting method is only suitable for KdV-type and NLS-type DSWs and is not valid for BO/BO-type DSWs. This realization prompted the development of a new approach, namely, by reversing the sign of dispersion in the associated BO system. He and his collaborators successfully adjusted the fitting method to make it generally applicable to BO systems. The modified DSW fitting method was then applied to the BO equation \cite{BOfitting}, the Calogero-Sutherland dispersive hydrodynamic system \cite{BOfitting}, and the Boussinesq-BO equation \cite{BBO}, yielding excellent agreement with full numerical solutions. 

While the mathematical theory for Dispersive Shock Waves (DSWs) in local media is well-established and developed, the same cannot be said for nonlocal media. Physical examples of nonlocal media include shallow waters with surface tension (capillary effect) \cite{markkaw, pat, patjump, patkawtrav, salehekdv}, polarized light waves in defocusing nematic liquid crystals \cite{nemboreori, nemboreel, salehnem1, salehnem2}, femtosecond pulses in nonlinear fiber optics \cite{yurifemto}, photorefractive crystals \cite{photobore} and thermal optical media \cite{thermal1, thermal2}.

A significant recent discovery is that DSWs in non-local media exhibit resonance phenomena. In this context, resonant radiation (or diffractive radiation, as referred to in optics) becomes in resonance with the individual waves of the DSW, leading to a fundamental alteration of the DSW's structure. This resonance effect, strongly influenced by the magnitude of initial discontinuity jumps and the values of higher-order dispersive coefficients, deviates from the classical form of DSWs, giving rise to novel and distinct types of DSW structures. The radical cause behind the existence of resonance in DSW structures lies in the non-convexity of the linear dispersion relations of the corresponding dispersive hydrodynamic systems \cite{markkaw, nemboreel}. Non-convex dispersion implies a profound interaction between waves with drastically different frequencies/wavenumbers.

Among these non-classical DSW types are Perturbed DSW (PDSW), Radiating DSW (RDSW), Cross-over DSW (CDSW), Travelling DSW (TDSW), and Vacuum DSW (VDSW). For detailed discussions and modulation theory solutions regarding these resonant DSWs, refer to \cite{nemboreel, markkaw, pat, salehnem1, salehnem2, salehekdv}. NFS and his collaborators made major contributions in studying these resonant DSWs, employing sophisticated methods based on modulation theory and perturbation theory to solve them. An intriguing and recently discovered modulation theory solution is Whitham Modulation Jump conditions, or so-called \textit{Whitham shocks}. These jump conditions can be viewed as the dispersive analog of Rankine-Hugoniot jump conditions for classical shocks but in dispersive systems. The concept of Whitham shocks has proven to be a powerful method for addressing numerous open questions, particularly those related to non-convex TDSW and CDSW resonant regimes. For detailed discussions of resonant DSWs influenced by non-convex dispersion and Whitham shock solutions for these classes of DSWs, see \cite{patjump, salehekdv}.

The field of ``non-convex dispersive hydrodynamics'' has now emerged as a prominent and actively researched area within nonlinear dispersive wave theory. This trend led to a very successful 6-month program generously hosted by the Isaac Newton Institute (INI) in 2022, entitled ``Dispersive Hydrodynamics: Mathematics, Simulation, and Experiments, with Applications in Nonlinear Waves.'' This program is followed by a sequel, a 1-month INI satellite event entitled ``Emergent Phenomena in Nonlinear Dispersive Waves,'' that is scheduled in 2024.

\section{\centering\sc\textcolor{purple}{Final Thoughts from students, collaborators and friends}}\label{s:thoughts1}

\begin{itemize}

\item \textcolor{purple}{Professor Roger Grimshaw, Loughborough University and University College of London}\\
I first met Noel at the end of 1984 when he came to the University of Melbourne to join me as a post-doctoral fellow.  He was my second research fellow, and his arrival was extremely positive for us both. When he came I had become interested in the newly observed ``upstream solitons'' especially in the internal waves context, and had derived a forced Korteweg-de Vries equation which I thought might be relevant. Noel came from Caltech where he had just finished a PhD and was armed with the Whitham modulation theory for nonlinear waves.  He immediately saw that this could be used to describe the ``upstream solitons,'' better called undular bores, and developed the modulated cnoidal wave theory, or solitary wave train theory, for their description.  The outcome was a JFM paper \cite{noelres1} which is by far my most cited paper, and I suspect is also Noel’s most cited paper. Modulated nonlinear waves and undular bores remained a theme  for Noel throughout his career, not only in fluids but in many other physical contexts. 

\quad In 1986 I moved to UNSW in Sydney and Noel followed me there. But he very soon took up his first academic position at the nearby University of Wollongong in 1987. Because of our common research interests and overlapping colleagues we remained in contact, mostly virtual. But after Noel had moved to Edinburgh and I had moved to Loughborough University in England an active collaboration was revived through my colleagues Gennady El and Yury Stepanyants. We even met up on several work-oriented occasions. 

\quad Noel will be greatly missed, not just by his family and friends, but by his many colleagues and the international community of nonlinear wave enthusiasts. He has passed from us at a relatively young age when he still had so much to offer.  My sincere condolences to his family, friends and colleagues. 

\item \textcolor{purple}{Professor Mark Hoefer, University of Colorado Boulder}

My career took a major positive turn upon meeting Noel over a dozen years ago.  Through the series of meetings \textit{Nonlinear Guided Waves} that he organized, Noel warmly and encouragingly welcomed me, a beginning junior researcher, into a vibrant community of nonlinear wave scientists with whom I had little previous exposure.  Noel was a driving organizer behind these meetings that took place in such exotic localities as Istanbul (Turkey), Santiago de Compostela (Spain), and Kingussie (Scotland).  His mentorship of junior colleagues was sincere and genuine, as exemplified by opening his home to stay during research visits by me and other junior colleagues.  It is a testament to his dedication and fortitude that, while he battled illness toward the end of his life, he continued to do research and mentor junior researchers.  I am indebted to Noel for his kindness and encouragement.  The nonlinear waves community has been enriched by Noel Smyth's contributions.

\item \textcolor{purple}{Professor Gennady El, Northumbria University} \\ 
I have been privileged to know Professor Noel Smyth for over two decades. Together we published six papers on undular bores, the subject for which we both shared great interest. Working with Noel has always been a great pleasure on both scientific and personal levels for me.  His kindness, honesty, quirky sense of humour, and the absence of even a slightest pretence, combined with the professionalism of the highest probe, made Noel a very special colleague and friend for me.   I have always felt great respect for his broad knowledge and deep insight into the subject of nonlinear waves, a distinct legacy of his PhD advisor, the great Gerald Whitham.  I am particularly grateful to Noel for introducing me to the \textit{Nonlinear Guided Waves Workshop} series, a very special club of nonlinear waves aficionados created by him and Gaetano Assanto with the basic principle of ``great science in great locations,'' that has left us with so many fond memories. 

\begin{figure}[!ht]
    \centering
\includegraphics[width=0.5\textwidth]{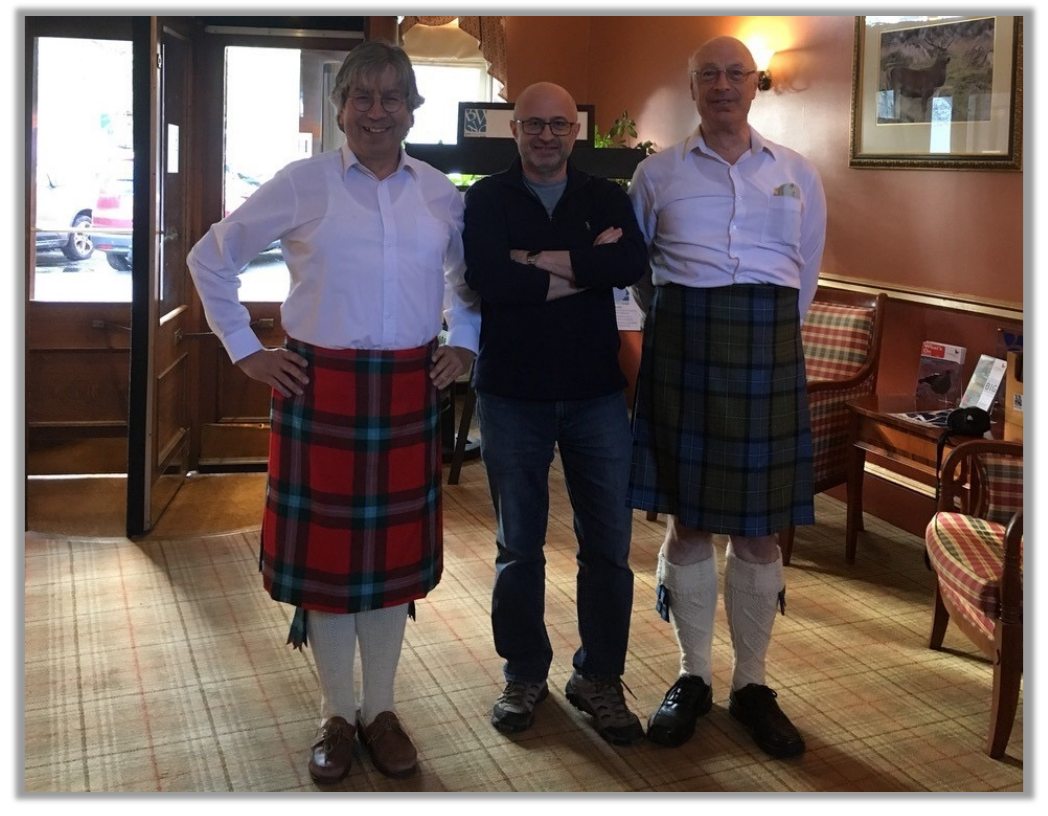}
    \caption{Grantown-on-Spey, during a \textit{Nonlinear Guided Waves Workshop}: Prof Gennady El between NSF and Gaetano Assanto wearing Scottish kilts.}
    \label{f:gennady}
\end{figure}

\item \textcolor{purple}{Professor Yuri Kivshar, Australian National University (ANU)} \\ 
I met Noel first time at the meeting on \textit{Nonlinear Coherent Structures} in Edinburgh. He was the only one wearing kilt at the banquet, and curiosity of that situation was that, as I was told, he was an Australian who got a job and moved to Scotland from Queensland. Very soon, some people introduced me to him, and we started our collaboration because we met again soon during one of his returns to Australia. 
    
\quad We have published 3 joint papers during 2007--2009 on vortex type beams in nematic liquid crystals, a topic pioneered by our common friend Gaetano Assanto.   Noel’s contribution to those papers was critical, his solid approach of a gifted applied mathematician helped to predict and analyse the stabilization of a vectorial vortex, when the vortex is unstable in the absence of the second beam. We also found  that the bright vortex can guide the beam in a stable manner, provided that the nonlocality is large enough.  Two of our common papers were cited more than 50 times each. 
I have interacted with Noel more closely during our remarkable trip to India in 2008, and almost all my photos of Noel come from that time. I share a few of those photos including a photo next to the Srinivasa Ramanujan statue.

\begin{figure}[!ht]
    \centering
\includegraphics[width=0.9\textwidth]{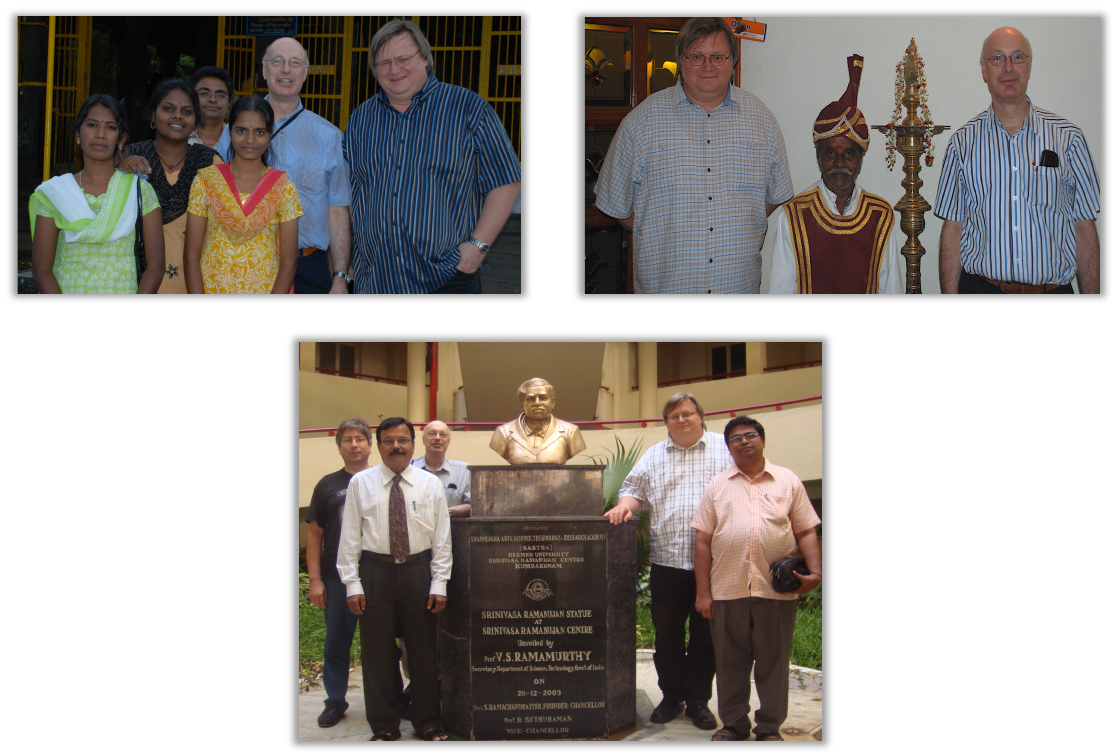}
    \caption{India, during a \textit{Nonlinear Guided Waves Workshop}: Prof Yuri Kivshar and NSF, together with colleagues.}
    \label{f:yuri}
\end{figure}

\item \textcolor{purple}{Professor Yury Stepanyants,  University of Southern Queensland} \\ 
With great grief, I learned about the death of my friend Noel. I was aware of his desperate struggle against a heavy illness within the last year but still hoped that it would recede. We met for the first time in 1994 at the conference in Edinburgh, where Noel was one of the organizers. For me, it was funny to observe him in the traditional Scottish attire, wearing a kilt.

\quad After that meeting, I started to follow Noel’s publications and found that his research interests were very close to mine. In 2017, I knew that Noel came to Australia to work during his vacation with our mutual friend Prof Timothy Marchant, who was at that time the Head of the Department at the University of Wollongong. Using this opportunity, I invited Noel to visit our University in Toowoomba to present a seminar and discuss problems of mutual interest. Noel accepted my invitation and told me that he would come by car, together with his mum, Gwen. I was surprised and argued that such a long trip, almost 1000 km, would be difficult for his mum and advised them to take a plane. But Noel calmed me down and said that they like to travel together by car. They knew well the route from Wollongong to Toowoomba and would take one stop halfway to spend one night in a motel. When Noel arrived with his mum in Toowoomba, I found them not overly tired and quite cheerful. We spent a few days together sightseeing in Toowoomba and its surroundings, dinners at restaurants and in my house. My wife, Irina, and I were charmed by their company. They were very open and shared their family stories with us.  


\quad Next year, in 2018, Noel with his mum visited us once again for a few days. And we, my wife and I, enjoyed their company. Every time, when visiting Toowoomba, Noel presented seminars at our School of Mathematics, and we had numerous discussions on scientific problems. The outcome of these visits was the publication of two papers jointly with Prof Roger Grimshaw.

\quad In 2019, I obtained a USQ grant for the invitation of Eminent Visiting Scholars. It was supposed that Noel would come to Toowoomba for a month for collaborative work with me and presentation of scientific and public seminars. Noel planned to come to Australia together with his son, Calum. He had booked an airplane ticket, but then he was forced to postpone the visit due to the COVID problem. We started to work distantly on the new problem and obtained some preliminary results, hoping that in the nearest years, the challenges posed by COVID would be overcome, and we would meet again in Toowoomba. But unexpectedly, Noel informed me that he was forced to stop working on the project due to a very serious problem with his health. We continued our communication by email, not touching scientific problems, and I believed that the medical treatment would help Noel to overcome the illness. Alas, in February 2023, I learned that he had passed away.

\quad We, my wife and I, will remember our remarkable friends Noel Smyth and his mum. We express our deep condolences to Noel’s family, his wife Juliet, son Calum, and relatives.

\item \textcolor{purple}{Professor William Kath, Northwestern University} \\
My professional collaboration with Noel began in earnest in 1992. I had known him since the Fall of 1980 when he arrived at Caltech for his graduate studies — I had arrived 2 years earlier — but we worked on different topics. By the time the early 1990s had rolled around, however, I had been at Northwestern for a few years working on pulse propagation in nonlinear optical fibers, while Noel was a well-known expert on nonlinear water waves and had moved to the University of Edinburgh. While at Edinburgh, Noel learned about the possibility of a collaborative research grant from NATO; he was incredibly good at identifying such opportunities. For some reason, he thought of me. After discussions and a joint back-and-forth writing exchange, he applied for both of us. Those familiar with such things will be surprised to know that when Noel was awarded the grant, NATO sent him the funds in cash, instructing him to deposit it in a new, separate, bank account so that the funds could accrue interest. Noel merely was required to account for the interest in the final report. Those NATO funds marked the official start of our collaboration. I visited Scotland for a month in the summer of 1992 so we could work together, and Noel visited Northwestern for an extended period in 1993.

\quad The next significant event in our joint professional interactions was in 1994. Professor David Parker had applied for and was awarded funds to host a \textit{European Study Foundation Study Centre on Nonlinear Optics and Guided Waves} in 1994. Noel was one of the local organizers, and he was able to get me invited to the gathering. This was one of the most memorable meetings that I ever attended. It ran for three weeks, and participants were a veritable “Who’s Who” of nonlinear optics and nonlinear waves, including applied mathematicians, applied physicists, and several luminaries from the world of fiber optics and light wave communications. I am incredibly grateful that Noel made such an opportunity possible for me.

\quad A related strong set of memories I have of Noel that may not seem like much to many people reading this is the use of the internet for research. Email started becoming much easier to use around the time Noel and I began collaborating, and I remember how wonderful it was the first time I sent a figure or comments about a draft of a paper we were writing to Noel at the end of the day in Chicago and had his response the next morning when I arrived at the office. The only way to really appreciate how much of an improvement this was would be to have lived and worked in the time when it was not available. Noel loved using such tools and was an earlier adopter of lots of different types of technology.

\quad All the above merely sets the stage for what I remember most and best about Noel, however. More than anyone else I have known, Noel was a master at blending work and fun, especially for those he was hosting. If you visited and worked with him, Noel made sure that you had a productive professional experience and that you truly enjoyed yourself during your free time, as well.

\quad This personality trait for Noel first showed itself when he was a graduate student. Noel’s Australia Day parties, for which he traditionally baked a pavlova, were legendary. And to give some idea of the vibe of these parties, I think it was the only time that I ever saw his advisor, Gerry Whitham, a bit tipsy.

\quad During both of our stays in Scotland, Noel used his formidable hosting skills to find housing for us. In 1992, this turned out to be a large, rambling house of a colleague who was away on sabbatical (a.k.a. study leave in the UK). In addition, Noel organized excursions on weekends to nearby sites or into the Scottish Highlands for my whole family. Noel escorted us on these trips and paid particular attention to doing things that would entertain my children, who were 3 and 5 at the time of our first trip. Seeing then how much he loved kids, we were not surprised later by how much he doted on his own son. I know he was immensely proud of Calum. Particularly memorable was a weekend trip to the Isle of Skye. On that trip, Noel’s conversation in Gaelic with the proprietress of the B\&B at which we were staying ensured that we were not treated as mere tourists, and we were invited to a céilidh — dining, drinking, and dancing at a local pub in the Scottish Highlands.

\quad Noel loved to travel and all the things that go with it. And while he lived for many years in Scotland and fully embraced his somewhat-removed Gaelic heritage, the Australian in him was never completely silenced. We saw this directly, because whenever he came to visit us in Chicago he would always want to go to dinner at Merle’s Barbecue, which opened in Evanston, Illinois, around the time that our collaboration began.

\quad Noel will be missed by my whole family, but we are grateful for the fond memories of our times with him.

\item \textcolor{purple}{Doctor Luke Sciberras, Lead Scientist at Panthera Finance} \\ 
Noel had an infectious curiosity for everything, not only the mathematical problems he was working on. Noel had infinite patience for trivial questions and always would lend a hand. My doctorate would not have been the same enjoyable experience without Noel. Noel passed on many life lessons during his mentorship, ones I am still using today. I am extremely lucky to have been able to call Noel a friend.

\item \textcolor{purple}{Doctor Panayotis Panayotaros, National Autonomous University of Mexico} \\ 
Noel visited UNAM in Mexico City for more than twenty-five years, and is greatly missed by the many people whom he met and interacted with there.

\quad He started visiting the Mathematics and Mechanics group of IIMAS, the Applied Math institute at UNAM, since at least the mid-1990s, initially to collaborate with the late Tim Minzoni, a fellow Whitham ex-alumnus who graduated from Caltech a few years before Noel. Noel and Tim collaborated on most of the main areas of Noel's research, including dispersive shocks (or, as Noel always insisted, undular bores), solitons, and radiation in many physical problems.

\quad Noel's friendly personality and interests in applied mathematics made him almost a member of the Mathematics and Mechanics group, and led to work with researchers in other scientific fields as well.

\quad It is accurate to say that Noel made a significant contribution to the academic life of UNAM for at least two decades, publishing papers, mentoring students and younger colleagues, organizing academic events, and more. His connection with closest collaborators at UNAM went further, up to common views of science, mathematics, and academia, and resulted in initiatives with a long-lasting positive impact on our university.

\begin{figure}[!ht]
    \centering
\includegraphics[width=0.5\textwidth]{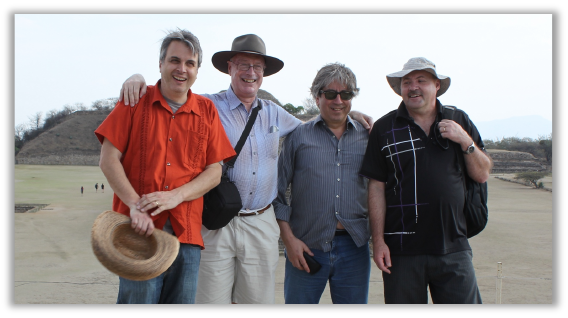}
    \caption{Oaxaca, during a \textit{Nonlinear Guided Waves Workshop}. From left to right: Dr Panayotis Panayotaros, NFS, Gaetano Assanto and Tim Marchant.}
    \label{f:gennady}
\end{figure}

\item \textcolor{purple}{Doctor Mike MacNeil, Senior Engineer at Apple} \\ 
I first met Prof Noel Smyth in person when he picked me up from the airport as I was starting my PhD. I remember being surprised by how tall he was, but he gave me a cheery ``Hello'' in his signature friendly, approachable style. We chatted small talk back and forth, he was always wondering “How is it going?” outside of science. But soon after that we started talking shop. It was then, maybe 15 minutes after landing in the UK I first found out about his towering intellect. He outlined his vision for a dispersive shock waves project he had in mind for my PhD, which he would talk about excitedly. I did not end up working on that project, but I remember thinking here is someone who is a scientist to the core. He absolutely loved nonlinear waves.

\quad This first meeting was indicative of how the rest of my PhD would go. He had a knack for mentorship, being a blunt but kind guiding hand while ensuring the bar was very high. If he considered part of your work “thin,” with patience he would recommend you go back to the drawing board to strengthen the argument, usually with a couple of keen insights of his own. That said, he did not mind cajoling his students. Early on in my PhD he laughed at my laptop as I showed him a quick result I was excited about. He said “You really should learn a real programming language at some point” (I was using MATLAB). I did, and to this day it is what I do professionally (although it is not Fortran, I am sorry Professor).

\quad He was very kind with his time. I remember the door of his office always being open, chalkboard ready and him smirking with a new idea. If it was not and he was in a meeting, he would always sit down with me afterward and help me out if I was stuck. Of course, it never took him long to figure out my mistake. 

\quad He was encouraging as well. Closer to the end of my PhD, I had an idea for an asymptotic method I was excited about. We would bounce ideas back and forth on it for hours. He would dive in, work alongside me, just as excited as I was, filling in missing pieces and pointing out where I had missed a spot. Encouraging me to present it as soon as possible. He was not just a great mentor, he felt like a great colleague.

\quad I think it is relatively uncommon to say this nowadays, but I loved doing my PhD. They were some of the best years of my life. One major reason is I got to spend time and learn from this clever, frank, friendly, tirelessly meticulous and outstanding Australian scientist. I am very proud, happy and downright lucky to have been a student of Prof Noel F. Smyth.

\item \textcolor{purple}{Doctor Patrick Sprenger, University of California Merced} \\ I spent six weeks in Edinburgh working with Noel during the summer of 2017. Prior to that, my interactions with Noel had been limited to email correspondence. My first in-person meeting with him took place at the Edinburgh Airport, where he kindly offered to pick me up upon my arrival. Noel's extraordinary generosity and kindness exceeded all reasonable expectations. The Smyth family graciously extended an invitation to stay in the guest bedroom of their home during my visit. Noel served as an exceptional mentor during my research stay. He went to great lengths to ensure that our collaboration received the highest priority. He rearranged his office and work schedule, resulting in a highly productive partnership. Our work together led to the characterization of novel dispersive shock waves, which later became a key focus of my PhD thesis. Noel's unwavering support was invaluable, and he was among the first to address me as ``Dr'' following the successful completion of my defense.

\quad I last saw Noel in December of 2022 when I was able to visit him and Juliet at their home. Despite his health challenges, Noel's enthusiasm for scientific discussions remained unwavering. He continued to mentor and showed a genuine interest in my latest projects. Noel will forever be remembered as a distinguished scientist and mathematician. His contributions to nonlinear wave modulation theory have left an indelible mark not only on my work but also on that of countless others. Beyond his scientific influence, I will always cherish Noel as the friend who warmly welcomed me into his home.

\item \textcolor{purple}{Doctor Benjamin Skuse, writer and editor} \\ Noel was my PhD supervisor in the late noughties. Looking back, I cannot have made a good first impression. I had no knowledge of coding, was only a middling mathematician, had no Master’s degree, had little to no background in physics – let alone the topic of my PhD, mathematical modelling of nonlinear optical solitons – and was more focused on fooling around making music than pursuing a serious academic career. If I were in his position, I would have sent me back to where I had come from in that very first meeting. But Noel did not do that.

\quad As I would discover in my four years under his mentorship, Noel had a bottomless well of kindness and patience only matched by his deep enthusiasm for and knowledge in nonlinear wave phenomena. Through countless conversations we had in his office – the door to which was always open to me – I slowly absorbed this knowledge and enthusiasm for the topic. But perhaps most important to my development as both an academic and a person was the trust and belief he placed in me, even when I did not believe in myself. Noel trusted me to present our work at international conferences, he trusted me to lecture in his stead while he was away (and house and dog-sit for him), and he even trusted me to complete my research and write up my thesis remotely when I decided on a whim that I wanted to move to France.

\quad His trust and belief not only allowed me to shed the imposter syndrome that haunts many a postgraduates so that I could complete my PhD, but also provided me with a self-belief I have retained ever since, giving me the confidence to pursue a completely different career outside academia as a science and technology writer. I am sorry I never said this to you during your lifetime, but thank you Noel for enriching my life.

\item \textcolor{purple}{Doctor Rosa-Maria Vargas, University of Bergen} \\ 
I had the privilege of meeting Prof Noel for the first time at the Universidad Nacional Autónoma de México in autumn 2014, where he was a close collaborator in the Department of Mathematics and Mechanics at IIMAS. Not only was he a respected researcher, but he was also a dear friend to my PhD supervisors. Approaching him as a student had a profound impact on me. Noel was a very special soul; he demonstrated joy in life, confidence in his work, deep knowledge in sciences, curiosity in history and traveling, infinite love for his family and pets and had a real taste for authentic Mexican food.

\quad Though this initial contact was brief, it was during the end of my PhD that I realized I wanted to establish a deeper collaboration with him in the future. This opportunity materialized in January 2019 when I embarked on a Postdoctoral research journey to Edinburgh - a city that became, from the very first instant, one of my favorite places in the world. The vibrant community at the University of Edinburgh and their beautiful buildings will remain forever in my memories. 

\quad Noel introduced me to the dispersive hydrodynamic world. We embarked on a project on dispersive shock waves on water waves. His invaluable guidance and collaborative spirit not only introduced me to new theoretical and numerical techniques but also enriched my knowledge in powerful programming languages. This experience undeniably broadened my research horizons and extended my network of collaborators, connecting me with former students and colleagues of Professor Noel. I must admit that our spoken communications were predominantly in Spanish; he enjoyed speaking in Spanish. I will never forget his ability to do mathematics with just a pen and tons of paper. I still have copies of the materials he shared with me, with outstanding and intricate computations. When it comes to Whitham Modulation Theory, Nonlinear Phenomena in Physics, and Scientific computing, you need a brilliant mind like Noel’s to succeed. He possessed clarity in his thoughts and knowledge, a profound understanding of physics, outstanding proficiency in sophisticate mathematical methods, and programming expertise in powerful computing languages. Beyond his intellect, Noel was an easygoing man, generous, always wearing a smile, and remarkably transparent in his thoughts and emotions, and he had an infinite love for his charming family: Juliet and Calum, whom I had the pleasure to meet, including
his pet, Toby.

\quad We have lost one of the most brilliant minds in Sciences, but we can be certain that his ideas and lines of research will continue to bear fruit for years to come. We all carry a piece of his wisdom and spirit with us, ensuring that his legacy endures in our collective work and contributions. His brilliance will continue to shine through his legacy and seminal ideas.

\item \textcolor{purple}{Doctor Cassandra Khan, University of Edinburgh}\\
I was very lucky to have Prof Noel Smyth as both my mentor and PhD advisor. He was one of those rare scientists who possessed both a tremendous intellect, coupled with unbounded curiosity, as well as a big heart and a generosity of spirit.

\quad Professionally, I found him to be a tireless, innovative, and enthusiastic researcher. He worked in so many capacities within the nonlinear waves community that I found myself surprised sometimes at the idea that he had time to rest or eat. I think that is a point worth emphasising: he was so very, very generous with his time, both at the University and in concert with researchers around the world.

\quad Personally, Prof Smyth was a devoted husband and father, with a zest for hiking which he indulged whenever his schedule would allow. I was surprised and delighted to learn about his deep and abiding love of Mexico, and all things Mexican (culinary, historical, architectural, musical, etc). When engaged, he would happily talk about any subject pertaining to Mexico, from Temple of Kukulkan at Chichen Itza, to the proper way to make mole sauce, to the diversity of spiritual practises, dress, and belief amongst the indigenous communities of Oaxaca. A particularly fond memory I have related to that is when I and his other advisees got together to host a birthday party for him at a good Mexican restaurant in town. It turned out to be a wonderful time: a large, boisterous, chatty party that went on for hours, with everyone laughing and talking late into the night.

\quad It is getting close to a year now since Prof Smyth passed, and I still feel his absence. I can say without hyperbole that I would not be the scientist I am today without his guidance and support; but I can now only pay my debt forward.

\item \textcolor{purple}{Doctor Bryan Tope, Former PhD student at  University of Edinburgh} \\ 
The first time I met Noel was trying to get on a master’s mathematics course. Noel was a lecturer at Edinburgh University and was organizing the course. I had newly arrived in Edinburgh from Australia and wanted to further my maths education. Noel wanted to know my mathematics background. I had an honors degree in Industrial Chemistry and was a little concerned about my maths education. I told him I did the first two years of the honors Pure Maths degree and studied applied maths in my second year. I said my studies were in a little-known university in NSW. He then mentioned he had lectured at the University of NSW. My exam results were good, and I was allowed to study the mathematics course.
    
\quad I was always amazed how much Noel and I had in common. Noel had studied for his PhD at Caltech. I did some travelling in the 1980’s and went to California. I found work in LA and did a lot of hill walking in the mountains close to Caltech. I even managed to play cricket on the Caltech grounds. He was interested in Australian military. My father was in the Australian army and my maternal grandfather had fought in WWI. He fought in Gallipoli, Passchendaele and was sent to Murmansk at the end of the war. I was amazed that Noel knew Australian troops had been sent to Murmansk at the end of the war. My paternal grandfather had gone to Dunkirk as the captain of a small ship.

\quad Noel had many interests. He was always willing to help students with their problems. If he could not help personally, he would put the students in contact with someone who could. He was always a pleasure to meet and helped people whenever he could. Noel also loved dogs. There always seemed to be a dog in his life. Sometimes you would come into his office and there would be a dog crouched under his desk.
    
\item \textcolor{purple}{Doctor Xin An, Cancer Counsil NSW} \\ 
I was very lucky to have Noel as my PhD supervisor. When I went to Edinburgh for my postdoc, he and his family selflessly took me in and let me stay with them for many weeks. Every evening he would cook for me, and after dinner, we would watch documentaries together till late. That was a great time. I admire and deeply respect Noel as a person and a researcher. I also extremely regret that there was not enough time to thank him for his generosity and guidance through the years.

\item \textcolor{purple}{Enrique Calisto, PhD student at University of Edinburgh} \\ 
When I think about Professor Noel, what always comes to mind first is his kindness and integrity. I will never forget that he and his family welcomed me, practically a stranger, into their home, when I was starting my PhD in Edinburgh. He showed me a lot of kindness and gave me much guidance and support. He had an unflagging enthusiasm and passion for science and was always ready to share his knowledge with patience. He set an example for me to continue working, and I will always be grateful for the encouragement he provided, even through the difficulties of his last year.

\item \textcolor{purple}{Doctor Khiem Nguyen, University of Glasgow} \\ 
I got to know Prof Smyth via emails in 2015 when I tried to look for a host for my internship. My PhD research topic was the modulation theory for nonlinear dispersive waves. I contacted three professors, and Prof Smyth was the only one who accepted to host me. Then, I wrote the research proposal to apply for the funding, and Noel helped me sharpened the proposal, which led to a successful application. I still could remember how carefully he gave feedback to my writing and proposal’s ideas, which had never happened to me before. For that moment, I knew I contacted the right person and got the proper support. I later learned that Noel was a PhD student of the celebrated Prof G.B. Witham.

\quad Noel Frederick Smyth, to me, was not only a collaborator but also a long-term advisor. I must admit that I learned about Whitham modulation theory from Noel more than from my own PhD advisor. He was always patient with all my technical and academic life related questions. When I sent him emails for questions, I mostly received replies very promptly within a day. After knowing Noel in 2016, he has become my life-long advisor. Whenever I have tough time at work or in my academic career choice, I always came back to ask him for advice. He was always straightforward and supportive in his answers and advice.

\quad Noel stands a true scientist and was always keeping high standards of working ethics. After finishing my research visit at Edinburgh and Philadelphia, I started my Postdoc position in Stuttgart, Germany. I explained to him that my employer was not happy with me collaborating with others outside the institute, and we could not continue collaboration after our first research paper. At that time, we were about to write a paper as a result from my research stay at Edinburgh. However, he misunderstood my explanation and said we did not have to submit the paper. He was happy that we could finish the problem we already started. For him, having a paper was not worth losing my job. He was always supportive to my academic career. On working on the first project, we slightly changed our target equations. For that reason, the project we had worked on initially was abandoned. Later, when I finally had time to complete the unfinished project, I could manage to inform Noel of our progress. He refused to be a co-author as he thought I worked on the project all alone. I had to remind him of the uncompleted project in 2016 when he made significant contributions. From these two occasions, I told Noel that he was a true scientist. And he humbly replied that was what he tried to be. Indeed, I learnt a lot from his working ethics and positive attitudes towards research academics.

\quad Noel was always a straightforward and easily approachable person. When he hosted me as a visitor, I invited him to my flat for dinner on the same day. He accepted without any doubt and asked whether he could bring his wife Juliet and son Calum. I felt so happy that his family was so easy going and friendly. He was very knowledgeable and talented at linguistics. He could speak several languages. He could communicate with my friends easily. He was frank and honest and certainly not afraid of delivering his opinion no matter the sensitivity of the topic. My friends and I could ask how many questions we wanted. My flatmates were impressed and liked him a lot.

\quad When I got the job offer from University of Glasgow, Noel was very happy for me. We finally met again in 2022 and had a nice and long talk the whole afternoon. Little did I know that was the last time I could see Noel in person.

\item \textcolor{purple}{Doctor Karima Khusnutdinova, Loughborough University} \\ 
It is very sad that we lost Noel Smyth so early. Noel possessed a remarkable breadth of research interests, and he was always willing to lend his expertise and support. In particular, he kindly served as an external examiner for some of my PhD students and generously contributed referee reports for numerous papers I sent to him. Noel always returned his referee reports within about a week, if not sooner, and he also immediately responded to all emails.  It was as though he never stopped working, which was remarkable! The paper we have contributed to this Special Issue carries Noel’s influence, and he will forever be remembered as a very kind and inspiring colleague.

\item \textcolor{purple}{Professor Alejandro Aceves, Southern Methodist University} \\ 
I first met Noel when I arrived as a graduate student at Caltech in 1981. Since I wanted to do research in nonlinear waves, I had the opportunity to learn much from him given that he was working on his dissertation with Whitham and that he was grading Whitham’s nonlinear waves homework when I took this course.

\quad Over the years, we did not keep in touch much although I would keep track of his work. It was not until the multiple trips that Noel did to Mexico, while visiting the Applied Mathematics Institute at the National University of Mexico (IIMAS-UNAM) that I had the opportunity to meet him again in person. Needless to say, his affection to my home country, Mexico, brought particular joy in me.

\quad Noel was a bright scientist; very able of the use of asymptotics and modulation theory that he applied to a wide range of problems in nonlinear wave.  More so, he was a genuinely  nice human being, with an ever present smile. We all miss him.

\item \textcolor{purple}{Professor James Hill, University of South Australia} \\ 
Noel Symth grew up and attended school in Brisbane, Queensland and attended the Toowong State High School, to become Dux of his year, receiving both the Mathematics and the Science School Prizes and topping the state in these subjects. In 1979, he graduated from the University of Queensland with a First Class Honours Degree, the University Medal and a prestigious PhD Scholarship from the California Institute of Technology. His degree from the University of Queensland combined with his course and PhD work at the California Institute of Technology provided Noel with the depth and the confidence to move forward as a mathematical researcher of the first rank.

\quad He was appointed Lecturer in the Department of Mathematics at the University of Wollongong late in 1986, and he took up the appointment early in 1987, remaining there for just over three years, leaving in 1990 to take up a Lectureship at the University of Edinburgh. On his arrival at the University of Wollongong he soon ignited activity and discussion and became a thoroughly reliable and enthusiastic colleague for collaboration. Noel became a key member in a small group of Applied Mathematicians including John Blake, Chris Coleman and myself working on mathematical modeling and novel problems associated with microwave heating. He was quickly successful in gaining an ARC Large Grant on the undular bore waves that occur in the Gulf of Carpentaria. While at the University of Wollongong, Noel attended the three ANZIAM conferences 1988-90, held at Leura, Ballarat and Coolangatta respectively.   

\quad In both his personal and professional lives, Noel exhibited considerable kindness and generosity towards others, as a research collaborator and as a sympathetic and understanding colleague. He was a very open person and for Noel a “spade” would always be a “spade”, and he believed there was considerable merit in always providing a frank and honest assessment of situations. He was certainly not afraid of giving his opinion, in a forthright manner. His colleagues very much appreciated and valued his generous and open disposition, and I am only one of the beneficiaries of that kindness amongst many others.

\quad Noel will be sorely missed in many parts of the world, and his colleagues extend their deepest condolences to his wife Juliet, son Calum and his other family.

\end{itemize}

\section{\centering\sc\textcolor{purple}{Final Thoughts from the family}}

Calum and I would like to thank all those who have coordinated the production of, and those who have contributed to, the Special Issue of the journal \textit{Wave Motion} and this tribute paper for Noel. Noel would have been very touched by the respect in which he was held by his academic colleagues.

Noel had a strong allegiance to Caltech and he was grateful for the academic opportunities which it had provided to him.

Noel had a love of dogs and had an interest in history. He was always there to give a helping hand when needed.

Noel was a proud Australian and he had intended to retire to Australia but, sadly, this was not to be.

\section{\centering\sc\textcolor{purple}{Acknowledgements}}
We are grateful to Karima Khusnutdinova for acting as the Associate Editor of \textit{Wave Motion} for this Special Issue. 
We thank NSF’s students, collaborators and friends for sharing their memories and thoughts about personal and professional interactions with him. We are also indebted to the staff of \textit{Wave Motion} for hosting this Special Issue. G. Assanto's work was supported by the Air Force Office of Scientific Research under award no FA8655-23-1-7026.


\begin{thebibliography}{99}

\bibitem{whitham}  G. B. Whitham. {\em Linear and nonlinear waves.} J. Wiley and Sons, New York (1974).

\bibitem{whitham1} G.B. Whitham, ``Non-linear dispersive waves,'' {\em Proc.\ Roy.\ Soc.\ London A,}{\bf 283,} 238–-261 (1965).

\bibitem{whithampert} G.B. Whitham, ``A general approach to linear and non-linear dispersive waves using a Lagrangian,'' {\em J.\ Fluid Mech.,}{\bf 22,} 273-–283 (1965).

\bibitem{whithamvar} G.B. Whitham, ``Variational methods and applications to water waves,'' {\em Proc.\ Roy.\ Soc.\ London A,}{\bf 299,} 6--25 (1967).

\bibitem{whithambio1} A.A. Minzoni and N.F. Smyth, ``Gerald Beresford Whitham, 13th December, 1927 --- 26th January, 2014,'' {\em Biographical Memoirs of Fellows of the Royal Society,} {\bf 61}, 555--577 (2015).

\bibitem{whithambio2} A.A. Minzoni and N.F. Smyth, ``Modulation theory, dispersive shock waves and Gerald Beresford Whitham,''{\em Physica D,} {\bf 333}, 6--10 (2016).

\bibitem{noelthesis} N.F. Smyth, ``Part I. Soliton on a Beach and Related Problems. Part II. Modulated Capillary Waves,'' Ph.D. thesis, California Institute of Technology (1984). 

\bibitem{henshaw} W.D. Henshaw, N.F. Smyth and D.W. Schwendeman, ``Numerical shock propagation using geometrical shock dynamics,'' {\em J.\ Fluid Mech.,} {\bf 171}, 519--545 (1986).

\bibitem{noelres1} R.H. Grimshaw and N.F. Smyth, ``Resonant flow of a stratified fluid over topography,'' {\em J.\ Fluid Mech.,} {\bf 169}, 429--464 (1986).

\bibitem{noelres2} N.F. Smyth, ``Modulation theory solution for resonant flow over topography,'' {\em Proc.\ R.\ Soc.\ A,}{\bf 409,} 79--97 (1987). 

\bibitem{ekdv}  T. R. Marchant and N.F. Smyth,  ``The extended Korteweg-de Vries equation and the resonant flow of a fluid over topography,'' {\em J.\ Fluid Mech.,} {\bf 221}, 263--288 (1990).

\bibitem{timkdvboundary} T. R. Marchant and N.F. Smyth, ``Initial-boundary value problems for the Korteweg-de Vries equation,'' {\em IMA journal of applied mathematics,} {\bf 47}, 247--264 (1991). 

\bibitem{minzonitribute} G. Assanto, G. Cruz, P. Panayotaros and N.F. Smyth, ``Antonmaria Alessio Minzoni: The work of the applied mathematician impacted mathematical physics and biophysics,'' {\em Physics Today} (https://doi.org/10.1063/PT.6.4o.20171009a). 




\bibitem{JSTQE} G. Assanto and N.F. Smyth, ``Light Induced Waveguides in Nematic Liquid Crystals,'' {\em J.\ Sel.\ Top.\ Quantum Electron.,} {\bf 22}, 4400306 (2016).

\bibitem{physdrev} G. Assanto and N.F. Smyth, ``Self-confined light waves in nematic liquid crystals,'' {\em Physica D,} {\bf 402}, 132182 (2020).

\bibitem{JNOPM} G. Assanto, A.A. Minzoni and N. F. Smyth, ``Light self-localization in nematic liquid crystals: modelling solitons in reorientational media,'' {\em J.\ Nonl.\ Opt.\ Phys.\ Mat.,} {\bf 18}, 657--691 (2009).

\bibitem{Macneil} J.M.L. MacNeil, N.F. Smyth and G. Assanto, ``Exact and Approximate Solutions for Optical Solitary Waves in Nematic Liquid Crystals,'' {\em  Physica D,} {\bf 284}, 1--15 (2014).

\bibitem{hinonl} A. Alberucci, C.P. Jisha, N.F. Smyth and G. Assanto, ``Spatial optical solitons in highly nonlocal media,'' {\em Phys.\ Rev.\ A,} {\bf 91}, 013841 (2015).

\bibitem{Minzoni} G. Assanto, A.A. Minzoni and N. F. Smyth, ``Light self-localization in nematic liquid crystals: modelling solitons in reorientational media,'' {\em J.\ Nonl.\ Opt.\ Phys.\ Mat.,} {\bf 18}, 657--691 (2009).

\bibitem{Panos} G. Assanto, P. Panayotaros and N. F. Smyth, ``Mechanical analogies for nonlinear light beams in nonlocal nematic liquid crystals,'' {\em J.\ Nonl.\ Opt.\ Phys.\ Mat.,} {\bf 27}, 1850046 (2018). 

\bibitem{2color} G. Assanto, N.F. Smyth and A.L. Worthy, ``Two color, nonlocal vector solitary waves with angular momentum in nematic liquid crystals,'' {\em Phys.\ Rev.\ A,} {\bf 78}, 013832 (2008).

\bibitem{3color}G. Assanto, K. Garcia-Reimbert, A.A. Minzoni, N.F. Smyth and A.L. Worthy, ``Lagrange Solution for Three Wavelength Soliton Clusters in Nematic Liquid Crystals,'' {\em Physica D,} {\bf 240}, 1213--1219 (2011).

\bibitem{antiguide} L. Marrucci, N.F. Smyth and G. Assanto, ``Optical vortices in antiguides,'' {\em Opt.\ Lett.,} {\bf 38}, 1618--1620 (2013).

\bibitem{vortexnematicon} G. Assanto, A.A. Minzoni and N. F. Smyth, ``Vortex confinement and bending with the aid of nonlocal solitons,'' {\em Opt.\ Lett.,} {\bf 39}, 509--512 (2014).

\bibitem{vectorvortex} G. Assanto, A.A. Minzoni and N.F. Smyth, ``Deflection of Nematicon-Vortex Vector Solitons in Liquid Crystals,'' {\em Phys.\ Rev.\ A,} {\bf 89}, 013827 (2014).

\bibitem{vortexvector} G. Assanto and N.F. Smyth, ``Soliton aided propagation and routing of vortex beams in nonlocal media,'' {\em J.\ Las.\ Opt.\ Photon.,} {\bf 1}, 105 (2014).

\bibitem{Yana} Y.V. Izdebskaya, W. Krolikowski, N.F. Smyth and G. Assanto, ``Vortex stabilization by means of spatial solitons in nonlocal media,'' {\em  J.\ Opt.,} {\bf 18}, 054006 (2016).

\bibitem{steering} G. Assanto, B. Skuse and N.F. Smyth, ``Solitary wave propagation and steering through light-induced refractive potentials,'' {\em Phys.\ Rev.\ A,} {\bf 81}, 063811 (2010).

\bibitem{refraction} G. Assanto, A.A. Minzoni, N.F. Smyth and A.L. Worthy, ``Refraction of Nonlinear Beams by Localised Refractive Index Changes in Nematic Liquid Crystals,'' {\em Phys.\ Rev.\ A,} {\bf 82}, 053843 (2010).

\bibitem{refra} G. Assanto, N.F. Smyth and W. Xia, ``Modulation Analysis of Nonlinear Beam Refraction at an Interface in Liquid Crystals,'' {\em Phys.\ Rev.\ A,} {\bf 84}, 033818 (2011).

\bibitem{scattering} A. Alberucci, G. Assanto, A.A. Minzoni and N.F. Smyth, ``Scattering of reorientational optical solitons at dielectric perturbations,'' {\em Phys.\ Rev.\ A,} {\bf 85}, 013804 (2012).

\bibitem{dark} G. Assanto, T. Marchant, A.A. Minzoni and N.F. Smyth, ``Reorientational versus Kerr dark and grey solitons using modulation theory,'' {\em Phys.\ Rev.\ E,} {\bf 84}, 066602 (2011).

\bibitem{coupleddark} G. Assanto, J.M.L. MacNeil and N.F. Smyth, ``Diffraction induced instability of coupled dark solitary waves,'' {\em Opt.\ Lett.,} {\bf 40}, 1771--1774 (2015).

\bibitem{PRL} G. Assanto, L.A. Cisneros, A.A. Minzoni, B. Skuse, N. F. Smyth and A.L. Worthy, ``Soliton steering by longitudinal modulation of the nonlinearity in waveguide arrays,'' {\em Phys.\ Rev.\ Lett.,} {\bf 104}, 053903 (2010).

\bibitem{temperature} G. Assanto, C. Khan, A. Piccardi and N.F. Smyth, ``Temperature control of nematicon trajectories,'' {\em Phys.\ Rev.\ E,} {\bf 100}, 062702 (2019).

\bibitem{volcano} G. Assanto, C. Khan and N.F. Smyth, ``Multi-hump thermo-reorientational solitary waves in nematic liquid crystals: Modulation theory solutions,'' {\em Phys.\ Rev.\ A,} {\bf 104}, 013526 (2021).

\bibitem{Sala} F. Sala, N.F. Smyth, U. Laudyn, M.A. Karpierz, A.A. Minzoni and G. Assanto, ``Bending reorientational solitons with modulated alignment,'' {\em J.\ Opt.\ Soc.\ Am.\ B,} {\bf 34}, 2459--2466 (2017).

\bibitem{bending} U. Laudyn, M. Kwasny, F. Sala, M.A. Karpierz, N.F. Smyth and G. Assanto, ``Curved solitons subject to transverse acceleration in reorientational soft matter,'' {\em Sci.\ Rep.,} {\bf 7}, 12385 (2017).

\bibitem{curving} U. Laudyn, M. Kwasny, M.A. Karpierz, N.F. Smyth and G. Assanto, ``Accelerated optical solitons in reorientational media with transverse invariance and longitudinally modulated birefringence,'' {\em Phys.\ Rev.\ A,} {\bf 98}, 023810 (2018).

\bibitem{spin} G. Assanto and N.F. Smyth, ``Spin-optical solitons in liquid crystals,'' {\em  Phys.\ Rev.\ A,} {\bf 102}, 033501 (2020), 

\bibitem{tunneling} G. Assanto, A.A. Minzoni, M. Peccianti and N.F. Smyth, ``Nematicons escaping a wide trapping potential: modulation theory,'' {\em Phys.\ Rev.\ A,} {\bf 79}, 033837 (2009).

\bibitem{extlight} L. Sciberras, A.A. Minzoni, N.F. Smyth and G. Assanto, ``Steering of optical solitary waves by coplanar low power beams in reorientational media,'' {\em J.\ Nonl.\ Opt.\ Phys.\ Mat.,} {\bf 23}, 1450045 (2014).

\bibitem{boundary} A. Alberucci, G. Assanto, D. Buccoliero, A.S. Desyatnikov, T.R. Marchant and N.F. Smyth, ``Modulation Analysis of Boundary Induced Motion of Nematicons,'' {\em Phys.\ Rev.\ A,} {\bf 79}, 043816 (2009).

\bibitem{OL22} E. Calisto, N.F. Smyth and G. Assanto, ``Optical isolation via direction-dependent soliton routing in birefringent soft-matter,'' {\em Opt.\ Lett.,} {\bf 47}, 459564 (2022).



\bibitem{elreview} G.A. El and M.A. Hoefer, ``Dispersive shock waves and modulation theory,'' 
{\em Phys. D,} \textbf{333}, 11--65 (2016).

\bibitem{nemboreori} N.F. Smyth, ``Dispersive shock waves in nematic liquid crystals,'' {\em Phys. D,} \textbf{333}, 301--309 (2016). 

\bibitem{nemboreel}  G. El and N.F. Smyth, ``Radiating dispersive shock waves in non-local optical media,'' {\em Proc.\ Roy.\ Soc.\ Lond.\ A,} {\bf 472}, 20150633 (2016).

\bibitem{markkaw}  P. Sprenger and M.A. Hoefer,``Shock waves in dispersive hydrodynamics with non-convex dispersion,'' {\em SIAM J. Appl.\ Math.,} {\bf 77}, 26--50 (2017).

\bibitem{pat}   M.A. Hoefer, N.F. Smyth and P. Sprenger, ``Modulation theory solution for nonlinearly
resonant, fifth-order Korteweg-de Vries, nonclassical, traveling dispersive shock waves,'' {\em Stud.\ 
Appl.\ Math.,} {\bf 142}, 219--240 (2019).

\bibitem{patjump}  P. Sprenger and M.A. Hoefer, ``Discontinuous shock solutions of the Whitham modulation equations and traveling wave solutions of higher order dispersive nonlinear wave equations,'' {\em Nonlinearity,} {\bf 33}, 3268--3302 (2020).

\bibitem{patkawtrav} P. Sprenger, T.J. Bridges and M. Shearer, ``Traveling Wave Solutions of the Kawahara Equation Joining Distinct Periodic Waves,'' {\em J.\ Nonl.\ Sci.,} {\bf 33,} 79 (2023).

\bibitem{salehekdv} S. Baqer and N.F. Smyth, ``Whitham shocks and resonant dispersive shock waves governed by the higher order Korteweg–de Vries equation,'' {\em Proc.\ R.\ Soc.\ A,}{\bf 479,} 20220580 (2023). 

\bibitem{hydraulic} N.F. Smyth and P.E. Holloway, ``Hydraulic jump and undular bore formation on a shelf break,'' {\em J.\ Phys.\ Ocean.,} {\bf 18}, 947–-962 (1988).

\bibitem{stratified} P.G. Baines, ``Topographic Effects in Stratified Flows,'' {\em Cambridge Monographs on Mechanics,} Cambridge (1995).

\bibitem{dambreak} J.G. Esler and J.D. Pearce, ``Dispersive dam-break and lock-exchange flows in a two-layer fluid,'' {\em J.\ Fluid Mech.,} {\bf 667,} 555-–585 (2011).

\bibitem{loweratmosphere} D.R. Christie, ``Long nonlinear waves in the lower atmosphere,'' {\em J.\ Atmos.\ Sci.,} {\bf 46}, 1462-–1491 (1989).

\bibitem{morning1} R.H. Clarke, R.K. Smith and D.G. Reid, ``The morning glory of the Gulf of Carpentaria: an atmospheric undular bore,'' {\em Monthly Weather Rev.},{\bf 109,} 1726–-1750 (1981).

\bibitem{morning2} V.A. Porter and N.F. Smyth, ``Modelling the Morning Glory of the Gulf of Carpentaria,'' {\em J.\ Fluid Mech.,}{\bf 454}, 1–-20 (2002).

\bibitem{nonlocaldsws} C. Barsi, W. Wan, C. Sun and J.W. Fleischer, ``Dispersive shock waves with nonlocal nonlinearity,'' {\em Opt.\ Lett.,}{\bf 32,} 2930-–2932 (2007).

\bibitem{superfluid} W. Wan, S. Jia and J.W. Fleischer, ``Dispersive superfluid-like shock waves in nonlinear optics,'' {\em Nature Phys.,} {\bf 3,} 46-–51 (2007).

\bibitem{photoref} G.A. El, A. Gammal, E.G. Khamis, R.A. Kraenkel and A.M. Kamchatnov, ``Theory of optical dispersive shock waves in photorefractive media,'' {\em Phys.\ Rev.\ A,}{\bf 76}, 053183 (2007).

\bibitem{fibers} G. Xu, A. Mussot, A. Kudlinski, S. Trillo, F. Copie and M. Conforti, ``Shock wave generation triggered by a weak background in optical fibres,'' {\em Opt.\ Lett.\,} {\bf 41,} 2656–-2659 (2016).

\bibitem{bec1} M.A. Hoefer, M.J. Ablowitz, I. Coddington, E.A. Cornell, P. Engels and V. Schweikhard, ``Dispersive and classical shock waves in Bose-Einstein condensates and gas dynamics,'' {\em Phys.\ Rev.\ A,}{\bf 74,} 023623.

\bibitem{bec2} G.A. El, A.M. Kamchatnov, V.V. Khodorovskii, E.S. Annibale and A. Gammal, ``Two dimensional supersonic nonlinear Schr\"{o}dinger flow past an extended obstacle,'' {\em Phys.\ Rev.\ E,}{\bf80,} 046317 (2009).

\bibitem{plasma1} D. Biskamp, ``Collisionless shock waves in plasmas,'' {\em Nucl.\ Fusion,} {\bf 13,} 719 (1973).

\bibitem{plasma2} R.Z. Sagdeev, ``The Fine Structure of a Shock Wave Front Propagated across a Magnetic Field in a Rarefied Plasma,''{\em
Sov.\ Phys.\ Tech.\ Phys.,}{\bf 6,} 867 (1962).

\bibitem{plasma3} N.J. Zabusky and M.D. Kruskal, ``Interaction of Solitons in a Collisionless Plasma and the Recurrence of Initial States,''{\em
Phys.\ Rev.\ Lett.}{\bf 15,} 240 (1965).

\bibitem{plasma4} R.J. Taylor, D. Baker and H. Ikezi, ``Observation of
Collisionless Electrostatic Shocks,'' {\em Phys.\ Rev.\ Lett.,}{\bf 24,} 206 (1970).

\bibitem{plasmapit} A.V. Gurevich and L.P. Pitaevskii, ``Nonstationary structure of a collisionless shock wave,'' {\em Sov.\ Phys.\ JETP,}{\bf  38,} 291 (1974).

\bibitem{analysis} H. Flaschka, M. Forest and D. McLaughlin. ``Multiphase averaging and the inverse spectral solution of the Korteweg de Vries equation,'' {\em Commun. Pure Appl. Math.,} {\bf 33,} 739–-784 (1980). 

\bibitem{fitting} G.A. El, ``Resolution of a shock in hyperbolic systems modified by weak dispersion,'' {\em Chaos,}{\bf 15,} 037103 (2005).

\bibitem{dswreview} G.A. El and M.A. Hoefer, ``Dispersive shock waves and modulation theory,'' {\em Phys. D,} {\bf 333,} 11-–65 (2016).

\bibitem{patkawahara} P. Sprenger and M.A. Hoefer, ``Shock waves in dispersive hydrodynamics with nonconvex dispersion,'' {\em SIAM J.\ Appl.\ Math.,}{\bf 77,} 26-–50 (2017).

\bibitem{salehnem1} S. Baqer and N.F. Smyth, ``Modulation theory and resonant regimes for dispersive shock waves in nematic liquid crystals,'' {\em Phys. D,}{\bf 403,} 132334 (2020). 

\bibitem{salehnem2} S. Baqer, D.J. Frantzeskakis, T.P. Horikis, C. Houdeville, T.R. Marchant and N.F. Smyth, ``Nematic
dispersive shock waves from nonlocal to local,'' {\em Appl.\ Sci.,}{\bf 11,} 4736 (2021). 


\bibitem{yurifemto} Y.S. Kivshar and G.P. Agrawal, {\em Optical Solitons. From Fibers to Photonic Crystals,} Academic Press, San Diego (2003).

\bibitem{thermal1} E.A. Kuznetsov and A.M. Rubenchik, ``Soliton stabilization in plasmas and hydrodynamics,'' {\em Phys.\ Rep.,}{\bf 142,} 103-–165 (1986).

\bibitem{thermal2} N. Ghofraniha, C. Conti, G. Ruocco and S. Trillo, ``Shocks in nonlocal media,'' {\em Phys.\ Rev.\ Lett.,}{\bf 99,} 043903 (2007).

\bibitem{photobore} M. Segev, B. Crosignani, A. Yariv and B. Fischer, ``Spatial solitons in photorefractive media,'' {\em Phys.\ Rev.\ Lett.,}{\bf 68,} 923-–926 (1992).

\bibitem{ekdhighermodu} T.R. Marchant and N.F. Smyth, ``An undular bore solution for the higher-order Korteweg–de Vries equation,'' {\em J. Phys. A: Mathematical and General,} {\bf 39,} 37 (2006). 
.
\bibitem{ekdvinteract} T.R. Marchant and N.F. Smyth, ``Soliton interaction for the extended Korteweg-de Vries equation,'' {\em IMA J. Appl. Math.,} {\bf 56,} 157--176 (1996). 

\bibitem{kdv1} T.R. Marchant and N.F. Smyth, ``The initial boundary problem for the Korteweg-de Vries equation on the negative quarter-plane,'' {\em Proc.\ Roy.\ Soc.\ A,}{\bf 458,} 857--871 (2002). 

\bibitem{kdv2} W.L. Kath and N.F. Smyth, ``Soliton evolution and radiation loss for the Korteweg–de Vries equation,''  {\em Phys.\ Rev.\ E}, {\bf 661}, (1995).

\bibitem{kathrad1} W.L. Kath and N.F. Smyth, ``Soliton evolution and radiation loss for the nonlinear Schr\"{o}dinger equation,'' {\em Phys.\ Rev.\ E}, {\bf 51}, 1484 (1995). 

\bibitem{kathrad2} W.L. Kath and N.F. Smyth, ``Radiative losses due to pulse interactions in birefringent nonlinear optical fibres,'' {\em Mathematical and Numerical Aspects of Wave Propagation,} edited by J.A. DeSanto, SIAM (1998). 

\bibitem{kathrad3} W.L. Kath and N.F. Smyth, ``Radiative losses due to pulse interactions in birefringent nonlinear optical fibres,'' {\em Phys.\ Rev.\ E}, {\bf 63}, 036614 (2001).  

\bibitem{kpnoel} A.A. Minzoni and N.F. Smyth, ``Evolution of lump solutions for the KP equation,'' {\em Wave Motion,} {\bf 24}, 291--305 (1996). 

\bibitem{mkdv} N.F. Smyth and A.L. Worthy, ``Solitary wave evolution for mKdV equations,''  {\em Wave Motion}, {\bf 21,} 263--275 (1995).

\bibitem{albalwi} M.D. Albalwi, T.R. Marchant and N.F. Smyth, ``Higher-order modulation theory for resonant flow over topography,'' {\em Phys. Fluids}, {\bf 29}, 7 (2017).

\bibitem{xin} Xin An, T.R. Marchant and N.F. Smyth, ``Dispersive shock waves governed by the Whitham equation and their stability,'' {\em Proc.\ Roy.\ Soc.\ A,} {\bf 474}, 20180278 (2018).

\bibitem{sine1} A.A. Minzoni, N.F. Smyth and A.L. Worthy, ``Evolution of two-dimensional standing and travelling breather solutions for the Sine–Gordon equation,'' {\em Phys. D,} {\bf 189}, 167--187 (2004). 

\bibitem{sine2} L.T.K. Nguyen, N.F. Smyth, ``Modulation Theory for Radially Symmetric Kink Waves Governed by a Multi-Dimensional Sine-Gordon Equation,'' {\em  J.\ Nonl. Sci.}, {\bf 33},  (2023).

\bibitem{sine3} A.A. Minzoni, N.F. Smyth and A.L. Worthy, ``Pulse evolution for a two-dimensional sine-Gordon equation,'' {\em Physica D,} {\bf 159}, 101--123 (2001).

\bibitem{sine4} N.F. Smyth and A.L. Worthy, ``Soliton evolution and radiation loss for the sine-Gordon equation,'' {\em Phys.\ Rev.\ E}, {\bf 60,} 2330 (1999).

\bibitem{sine5} A.A. Minzoni and N.F. Smyth, ``A modulation solution of the signalling problem for the equation of self-induced transparency in the Sine–Gordon limit,'' {\em Meth. Applic. Analysis}, {\bf 4,} 1--10 (1997). 

\bibitem{modulnematic} A.A. Minzoni, N.F. Smyth and A.L. Worthy, ``Modulation solutions for nematicon propagation in nonlocal liquid crystals,'' {\em J.\ Opt.\ Soc.\ Am.\ B}, {\bf 24}, 1549--1556 (2007)

\bibitem{mdpinembook} G. Assanto and N.F. Smyth, {\em Light Beams in Liquid Crystals.} Multidisciplinary Digital Publishing Institute-Basel (Switzerland, 2022).

\bibitem{excellent} A. Alberucci, G. Assanto, J.M.L. MacNeil and N.F. Smyth, ``Nematic liquid crystals: An excellent playground for nonlocal nonlinear light localization in soft matter,'' {\em J. Nonl. Opt. Phys. Mat.}, {\bf 23}, 1450046 (2014). 

\bibitem{colloid1} X. An, T.R. Marchant and N.F. Smyth ``Optical dispersive shock waves in defocusing colloidal media,'' {\em Physica D}, {\bf 342}, 45--56 (2017).

\bibitem{colloid2} T.R. Marchant and N.F. Smyth, ``Semi-analytical solutions for dispersive shock waves in colloidal media,'' {\em J.\ Phys.\ B: Atom.\ Mol.\ Opt.\ Phys.,} {\bf 45,} 145401 (2012).  

\bibitem{colloid3} T.R. Marchant and N.F. Smyth, ``Solitary waves and their stability in colloidal media: semi-analytical solutions,'' {\em Dyn. Continuous, Discrete and Impulsive Systems Series B: Applic. Algor.,} {\bf 19}, 525--541 (2012). 

\bibitem{whithbous} R.M. Vargas-Magaña, T.R. Marchant and N.F. Smyth, ``Numerical and analytical study of undular bores governed by the full water wave equations and bidirectional Whitham–Boussinesq equations,'' {\em Phys. Fluids,} {\bf 33,} 067105 (2021). 

\bibitem{BOfitting} G.A. El, L.T.K. Nguyen and N.F. Smyth, ``Dispersive shock waves in systems with nonlocal dispersion of Benjamin–Ono type,'' {\em Nonlinearity}, {\bf 31}, 1392 (2018). 

\bibitem{BBO} L.T.K. Nguyen and N.F. Smyth, ``Dispersive shock waves for the Boussinesq Benjamin–Ono equation,''{\em Studies Appl. Math.}, {\bf 147,} 32--59 (2021).

\bibitem{fullwater} G.A. El, R.H.J. Grimshaw and N.F. Smyth, ``Unsteady undular bores in fully nonlinear shallow-water theory,'' {\em Phys. Fluids}, {\bf 18} 2423--2435 (2006). 

\bibitem{camassa} T.R. Marchant and N.F. Smyth, ``Undular bore solution of the Camassa-Holm equation,'' {\em Phys. Rev.\ E}, {\bf 73,} 057602 (2006).

\bibitem{magma} T.R. Marchant and N.F. Smyth, ``Approximate solutions for magmon propagation from a reservoir,'' {\em IMA J. Appl. Math.,} {\bf 70,} 796--813 (2005). 










\end{thebibliography}
\end{document}